# Giant entropy change at the co-occurrence of structural and magnetic transitions in the Ni$_{2.19}$Mn$_{0.81}$Ga Heusler alloy

L. Pareti[1], M. Solzi[2], F. Albertini[1], A. Paoluzi[1]

[1] IMEM - CNR, Parco Area delle Scienze, I-43010 Fontanini, Parma, Italy

[2] INFM and Dipartimento di Fisica, Università di Parma, Parco Area delle Scienze, I-43100, Parma, Italy

**Abstract**

We have studied the isothermal entropy change around a first-order structural transformation and in correspondence to the second-order Curie transition in the ferromagnetic *Heusler* alloy Ni$_{2.15}$Mn$_{0.75}$Ga. The results have been compared with those obtained for the composition Ni$_{2.19}$Mn$_{0.81}$Ga, in which the martensitic structural transformation and the magnetic transition occur simultaneously. With a magnetic field span from 0 to 1.6 T, the magnetic entropy change reaches the value of 20 J/kg K when transitions are co-occurring, while 5 J/kg K is found when the only structural transition occurs.

Adiabatic temperature changes and isothermal magnetic entropy variations are manifestations of the magnetocaloric effect (MCE), which is a fundamental property of magnetic materials. MCE is at the basis of the magnetic refrigeration process, which represents a promising alternative to the conventional gas compression technique in domestic and industrial refrigeration, because it is environmentally-friend and energy-saving. Large efforts are being done in order to individuate new classes of materials that fulfil the main requirements for the applications, i.e. high MCE around RT, obtained by the application of magnetic fields that can be produced by permanent magnets (lower than 2 T). Up to recently the research has been practically performed on rare-earth elements or compounds, because MCE, besides to be proportional to the temperature variation of the magnetisation and to the maximum applied field, it also depends on the size of the atomic magnetic moment [1]. Large values of the magnetic entropy change (defined *giant*) have been indeed found in RE-based compounds like $Gd_5(SiGe)_4$, $Tb_5(SiGe)_4$ [2, 3, 4]; however, high MCE have also been recently reported on transition-metals-based compounds like (MnFe)(PAs) [5]. A common feature of these new classes of magnetocaloric materials is that they undergo a concomitant first-order structural and magnetic transition, thus giving rise to giant isothermal entropy changes, when cycling around transition temperatures [4, 6].

Ferromagnetic martensites, and particularly $Ni_2MnGa$ *Heusler* alloys, which pertain to the class of *shape memory* materials, have recently drawn attention because of the possibility to induce huge strains (of the order of a few percent) by the application of magnetic fields [7, 8, 9, 10]. These compounds undergo a martensitic transformation between a low temperature tetragonal phase (which is magnetically hard) and a high temperature cubic one (magnetically soft) [11, 12]. This difference in the anisotropy strongly modifies the field dependence of the magnetisation in the two phases, with the saturation magnetisation value being slightly lower in the cubic austenite [3, 13]. Some recent works have evidenced the occurrence of significant isothermal variations of the magnetic entropy in NiMnGa compounds (up to $|\Delta S_m|$ =18 J/kg K for $\mu_0 \Delta H$ = 5 T) in correspondence to the martensitic transformation. In these cases, the martensitic transition temperatures $T_m$ ($T_{ma}$ on heating and $T_{am}$ on cooling) are lower than the Curie temperature $T_c$ and, as a consequence, the martensitic transformation takes place between two ferromagnetic phases [3, 13, 14].

On the other hand, both $T_c$ and $T_m$ can be widely tuned by varying the composition (x, y, z) of the alloy $Ni_{2+x}Mn_{1+y}Ga_{1+z}$, with x+y+z = 0. It has been shown that, for particular compositions, it is possible to obtain the simultaneous occurrence of the first-order structural martensitic transformation and of the Curie transition [15]. On the basis of the actual knowledge, this fact should realise the right conditions for the occurrence of large magnetocaloric effects.

The aim of the present work was to prepare a NiMnGa sample of modified composition in order to achieve a concomitant first-order structural and magnetic transition from the ferromagnetic martensite to the paramagnetic austenite and to study the magnetocaloric effect around such a transformation. There are various compositional possibilities [15, 16] for the simultaneous occurrence of $T_m$ and $T_c$. In the present work this has been obtained by increasing Ni content at expenses of Mn.

In order to compare the amplitude of the MCE occurring at the simultaneous structural/magnetic transition with that detectable as a consequence of the only first-order martensitic transformation, an alloy having distinct transitions, with $T_m < T_c$, has also been made.

Polycrystalline samples of composition $Ni_{2.19}Mn_{0.81}Ga$ (for the simultaneous occurrence of the transitions, henceforth called Ni19) and $Ni_{2.15}Mn_{0.75}Ga$ (henceforth Ni15) were prepared by arc melting technique. In order to improve both the stoichiometric homogeneity and the degree of lattice order, a heat treatment at 845°C for 72 hours was performed. The monophasicity of the samples was proved by X-ray powder diffraction and thermomagnetic analysis (TMA). The latter technique, which consists in the measure of the temperature dependence of the initial a.c. susceptibility, was also used to determine both Curie and martensitic transformation temperatures. A high sensitivity, stationary pendulum magnetometer was used to measure isotherm $M(H)$, as well as iso-field $M(T)$ magnetisation curves.

The results of TMA measurements, performed in an applied magnetic field of 0.5 mT, are reported in Figs. 1 and 2 for the samples Ni15 and Ni19 respectively. In the sample Ni15, the martensite/austenite transformation temperature $T_{ma} = 310$ K, the value corresponding to the inflection in the $\chi(T)$ drop, and $T_c = 335$ K are well separated. No thermal hysteresis is observed at the second-order Curie magnetic transition, while a little hysteresis separates the martensite/austenite from the reverse austenite/martensite first-order transformation, occurring on cooling ($T_{am}$ = 307 K). On the other hand, in the sample Ni19, a single hysteretic transition (with $\Delta T$ = 8 K) is observed between the ferromagnetic and paramagnetic states. The ordering temperature is thus 365 K on heating, corresponding to $T_{ma}$, while it is 357 K on cooling ($T_{am}$). The magnetisation curves M(T), performed at different values of the applied field, confirm that the magnetic transition is driven by the structural transformation (Fig.3). In fact the ferromagnetic/paramagnetic transition remains sharp even upon the application of large magnetic fields, which instead gives rise to a broadening of the magnetisation decrease at a normal second-order Curie transition. The origin of the two small kinks at both sides of the main sharp transition is explained below. It is noticeable that the transition temperatures are only slightly affected by the application of magnetic fields in Ni19, contrary to that reported in [16] for the parent alloy composition $Ni_{2.12}MnGa_{0.88}$.

From X-ray diffraction and TMA, both samples result to be homogeneous and single phase. However, the subdivision of a Ni19 sample of 50 mg (whose TMA is reported in Fig. 2) into smaller pieces, gives the type of TMA shown in Fig. 4. It is evident that the main part of the specimen shows unchanged values of transition temperatures, while a distribution of smaller and larger values occurs. The reported behaviour is usually attributed to a multi-phase non-homogeneous sample. Nevertheless, in this case, where a stoichiometric inhomogeneity was not present in the original sample, the peculiarity of TMA has to be ascribed to the possibility that structural strains, occurring as a consequence of the subdivision, modify the transformation temperature in parts of the sample volume. It is worth noticing that, in spite of the occurrence of different transition temperatures, the thermal hysteresis is still present with unchanged amplitude, indicating that $T_{ma}$ ($T_{am}$) and Curie temperature are tightly connected in this sample. Because of the high sensitivity of the available magnetometer, M(H) and M(T) curves have been performed using, as sample Ni19, the one giving the TMA of Fig. 4. No modification of the TMA has been observed upon subdivision in the Ni15 sample.

In order to evaluate the MCE in Ni15 and Ni19 samples, the indirect technique, consisting in the measurements of M(H) isotherms at different temperatures and the calculation of $\Delta S_m(T, \Delta H)$ by means of Maxwell equations, was used. The expression for the isothermal-isobaric magnetic entropy change is [17]

$$\Delta S_m(T, \Delta H) = \int_{H_i}^{H_f} \left( \frac{\partial M(T,H)}{\partial T} \right)_H dH \qquad (1)$$

where $H_i$ and $H_f$ are the initial and final values of the applied magnetic field, with $\Delta H = H_f - H_i$.

Some doubts have recently been raised on a possible overestimation of the isothermal magnetic entropy change, when using Maxwell equations in correspondence to first-order transitions [18]. Arguments have been brought against this hypothesis [19, 20] but the question does not seem to be yet clarified. In spite of that we present our data that can be compared with most of the literature values obtained in the same way on various materials [1, 3, 4, 13].

Isothermal magnetisation curves, performed up to 1.6 T, with increasing field (in steps of $\mu_0 \delta H \approx 80$ mT), are reported in Fig. 5 for the two samples. Measurements have been performed both above and below the transition temperatures. Details on the temperature steps are given in the figure.

The values of the magnetic entropy changes, calculated by a numerical integration of the expression (1) over a field span $\mu_0 \Delta H = 1.6$ T, are reported in Fig. 6 as a function of the average temperature between two next M(H) isotherms. Following the indications given in [21], an evaluation was done of the inaccuracy in the derivation of $\Delta S_m$, which accounts for errors in H, T and M measure. The combined error is represented, in the figure, by vertical bars.

A sharp peak, corresponding to the first-order transformation and a broader one in coincidence with the Curie transition, are observed in the Ni15 sample. The values of $\Delta S_m$ at $T_{ma}$ and at $T_c$ compare well with literature data [14].

The effect of the concentration of the two transitions in a narrow temperature span is evident from the data of Ni19 (Fig. 6b), where a maximum $|\Delta S_m|$ value of 20 J/ kg K is found. This value is about four times larger than that observed when the only structural transition occurs. The field span dependence of the peak value of $\Delta S_m$ was found to be linear. In the case of Ni19, the presence of two lateral peaks is the consequence of different thermomagnetic behaviours in distinct sample regions, as deduced from TMA (Fig. 4).

By performing a numerical integration to evaluate the area subtended by the peaks in $\Delta S_m(T)$, the magnetic refrigerant capacity, defined as:

$$q = \int_{T_{cold}}^{T_{hot}} \Delta S(T, P, \Delta H)_{P, \Delta H} dT \qquad (2)$$

can be obtained. $T_{cold}$ and $T_{hot}$ are the extrema of the considered temperature interval. The quantity $q$ represents the amount of heat that can be ideally transferred from a cold to a hot sink [22]. Performing the integration over the same $\Delta T$ range (50 K, with $T_{cold}$ = 300 and 340 K, $T_{hot}$ = 350 and 390 K, for Ni15 and Ni19 respectively) the values of 55 J/Kg and 49 J/Kg are found for $q$ in Ni15 and Ni19 respectively. In terms of volumetric units, the magnetic refrigerant capacity $q$ is 454 (404) J/m$^3$ for Ni19 (Ni15). For this calculation the density value of 8.25 g/cm$^3$ was deduced from X-ray diffraction measurements of the tetragonal phase in Ni19. From the similarity of the $q$ values in the two alloys, the summability of the isothermal entropy change for the refrigerant capacity could be inferred.

The obtained values for the maximum $\Delta S_m$ and for the magnetic refrigeration capacity $q$, represent excellent features, particularly because they have been reached with the application of a relatively low maximum magnetic field (1.6 T), which is attainable with a special configuration of permanent magnets [23]. Similarly to the case of $Gd_5Si_{1.8}Ge_{2.2}$, the large MCE value is connected to a concomitant first-order magnetic (order/disorder) and crystallographic (order/order) phase transition. In Ni19 alloy the high MCE effect is made possible by the large Mn magnetic moment (3.4 $\mu_B$, if deduced from RT magnetisation data in the stoichiometric $Ni_2MnGa$ alloy [9]) and realised through the rapid variation of the magnetisation around the firs-order magnetic transition. It is confirmed that the effective action of a first-order transition is to confine the occurrence of MCE in a narrow temperature range, thus enhancing its maximum value.

A further interesting characteristic of NiMnGa-based materials is that the observed MCE is reversible. In addition, these alloys are low cost and easy to prepare. These features make them attractive for applications in the field of commercial magnetic refrigeration. However, for this purpose, it is important to have a large MCE around room temperature. Thus, the effects of partial substitutions of the components will be explored with the aim of lowering the temperature of the martensitic transformation, while keeping its co-occurrence with the magnetic transition. Furthermore, being isothermal entropy changes a necessary, not sufficient condition for high adiabatic temperature changes, a direct measurement of $\Delta T_{ad}$ around the transition will be performed.

**Figure captions**

Figure 1. Temperature dependence of the initial a.c. susceptibility, for the sample $Ni_{2.15}Mn_{0.75}Ga$, under an applied magnetic field of 0.5 mT. Arrows indicate increasing and decreasing temperature branches.

Figure 2. Temperature dependence of the initial a.c. susceptibility, for the sample $Ni_{2.19}Mn_{0.81}Ga$, under an applied magnetic field of 0.5 mT. Arrows indicate increasing and decreasing temperature branches.

Figure 3. Temperature dependence of the specific magnetisation M measured at two different applied magnetic field values in $Ni_{2.19}Mn_{0.81}Ga$. (open symbols) $\mu_0 H = 0.2$ T; (filled symbols) $\mu_0 H = 1.5$ T. Arrows indicate increasing and decreasing temperature branches. Lines are guides for eyes.

Figure 4. Temperature dependence of the initial a.c. susceptibility, for a part (about 10 mg) of a 50 mg $Ni_{2.19}Mn_{0.81}Ga$ sample (the susceptibility of the whole specimen is reported in Fig. 2). Arrows indicate increasing and decreasing temperature branches.

Figure 5. Isothermal magnetisation curves at various temperatures for: (a) $Ni_{2.15}Mn_{0.75}Ga$ and (b) $Ni_{2.19}Mn_{0.81}Ga$, obtained with a magnetic field step $\mu_0 \delta H \approx 80$ mT.

Figure 6. Magnetic entropy change $\Delta S_m$ as a function of the average temperature between two next magnetic isotherms, with a field span from 0 to 1.6 T, for: (a) $Ni_{2.15}Mn_{0.75}Ga$ and (b) $Ni_{2.19}Mn_{0.81}Ga$.

**Fig. 1**

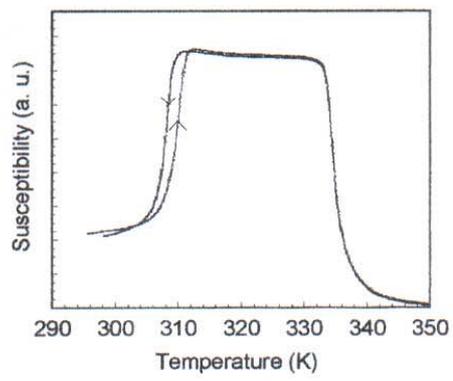

**Fig. 2**

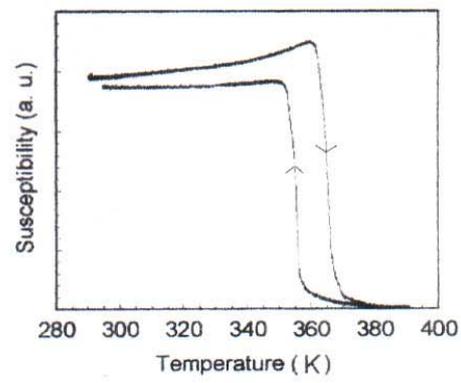

**Fig. 3**

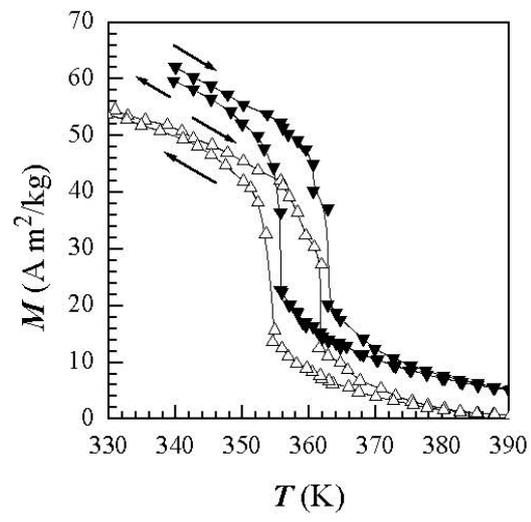

**Fig. 4**

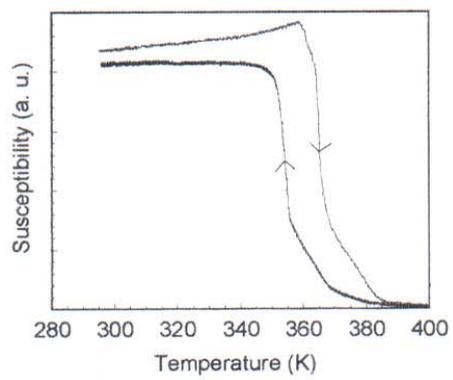

**Fig. 5**

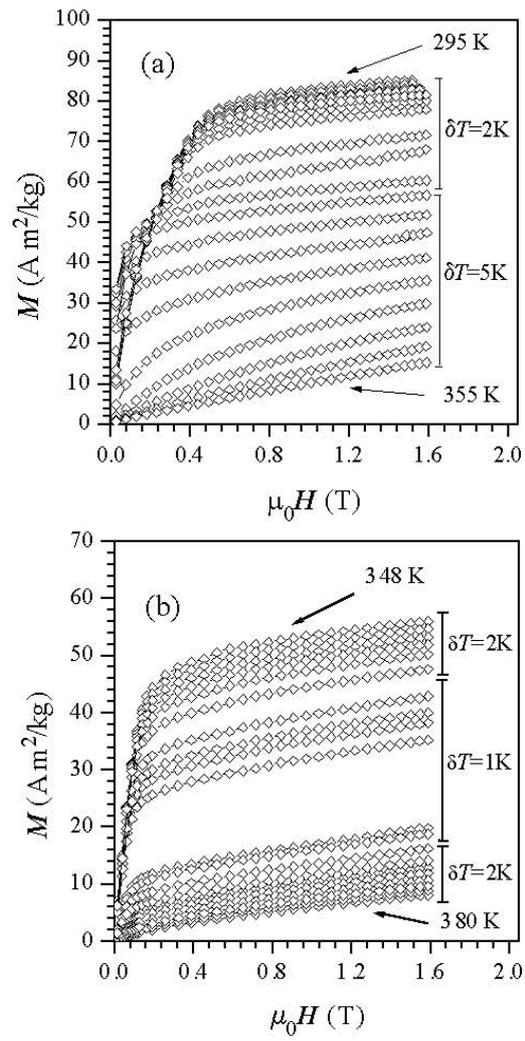

**Fig. 6**

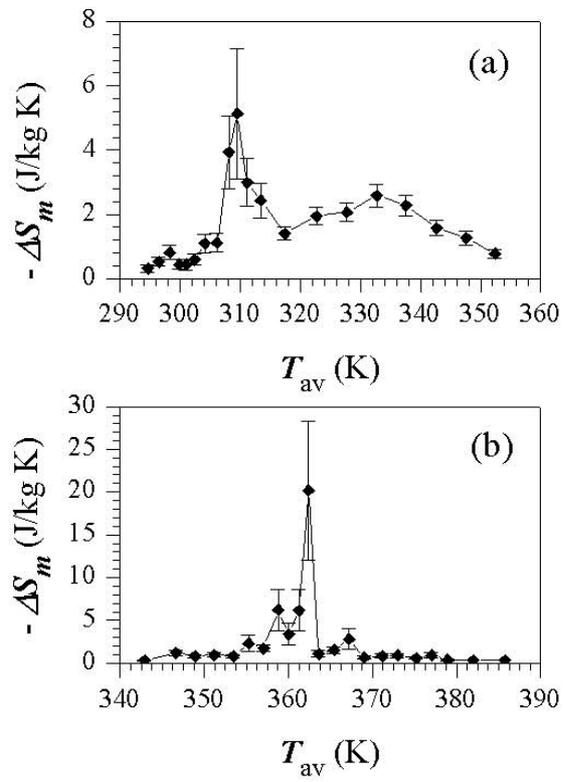